\documentstyle[aps,epsfig]{revtex}
\begin{document}
\draft
\title{Effects of mechanical strain on thermal denaturation of DNA} 
\author{Joseph Rudnick and Robijn Bruinsma\cite{current}}
\address{Department of Physics and Astronomy, UCLA, Box 951547, Los Angeles, 
California 90095-1547}
\date{\today}
\maketitle
\begin{abstract}
A class of simple statistical mechanical models for DNA melting, first 
proposed by Poland and Scheraga, have been demonstrated to exhibit 
first or second order thermodynamic singularity, notwithstanding the 
intrinsic one-dimensional nature of the problem.  In the present paper 
we show that the inclusion of twist elastic energy in the 
Poland-Scheraga models either leads to suppression of the 
thermodynamic singularity or to a weak, third order singularity.
\end{abstract}


The thermodynamic properties of DNA near the melting, or denaturation, 
point has become the focus of intense interest in the theoretical 
physics community \cite{PS,mukamel,BP,hwa,L&N}.  A class of simple 
models, proposed in their original form by Poland and Scheraga (PS) 
\cite{PS} exhibits a thermodynamic singularity, notwithstanding the 
essentially one-dimensional nature of the DNA molecule.  In these PS 
models the single-stranded DNA sections inside denaturation bubbles 
are treated as excluded-volume polymers performing a random walk.  
This produces an effective long-range interaction that leads to a 
thermodynamic singularity in this one-dimensional system.  Depending 
on whether excluded volume effects between single and duplexed strands 
are or are not taken into account, the thermodynamic singularity can 
have the character of either a first \cite{mukamel} or a second-order 
\cite{BP} phase transition.  We denote those two alternatives as Case 
I and Case II, respectively.

\begin{figure}
\centerline{\epsfig{file=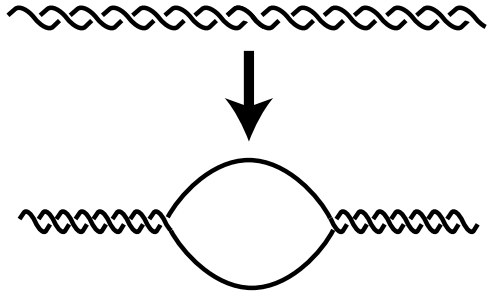,height=1.5in}}
\caption{The creation of a denaturation bubble induces extra twist in 
the undenatured portions of DNA. }
\label{fig:loops}
\end{figure}

Experimental studies of DNA melting have not yet provided any clear 
support for the presence of a thermodynamic singularity.  In 
particular, optical absorption studies of the transition usually yield 
smoothly sigmoidal behavior \cite{GBL}, although this might be 
attributed to finite size effects or to the sequence heterogeneity of 
DNA. However, the PS models do not incorporate an additional important 
effect, namely \emph{elastic strain}.  As shown in Figure 
\ref{fig:loops}, because of the interwinding of the two strands of 
duplexed DNA a denaturation bubble induces extra elastic twist in the 
undenatured portions of the strand.  For circular DNA, this strain 
energy cannot be relieved in the absence of special cutting enzymes, 
and progessive supercoiling can be expected to arrest the denaturing 
transition.  The thermal denaturation of circular DNA is 
well-known to be sensitive to the degree of over- or underwinding of 
the loop before it is closed \cite{B&B}.  In the biophysics literature, 
sigmoidal thermodynamic behavior is also connected with the 
development of supercoiling \cite{KLT}.

It is the aim of this letter to demonstrate that when supercoling is 
included in the PS model, the thermodynamic singularity is, indeed, 
smeared out or seriously weakened.  The starting point is the 
inclusion of distributed twist energy \cite{KLT} in the PS 
Hamiltonian:
\begin{equation}
{\cal H} = K\frac{\left(N_{2}-\sigma(N_{2}+N_{1}) \right)^{2}}{N_{1}} 
- \epsilon_{0} N_{1} + U_{\rm EV}
	\label{newham}
\end{equation}
The first term in (\ref{newham}) is the distributed twist elastic 
energy.  The untwisting imposed by the unbinding of $N_{2}$ base pairs 
is assumed to produce a twist proportional to $N_{2}$ on the remaining 
$N_{1}$ intact bases.  Here, $\sigma$ quantifies the extent to which 
the circular chain is underwound or overwound.  The limit $\sigma=1$ 
corresponds to complete unwinding of the DNA. When $\sigma <0$, the 
DNA duplex is overwound.  The constant $K$ is proportional to the 
torsional rigidity of double helical DNA ($ K \sim 10^{-12}$ erg).  
The second and third terms describe, respectively, the base-pairing 
energy ($\epsilon_{0} \sim 2$ kcal/mole) and the excluded volume 
interaction mentioned above.

We will examine the phase behavior of this Generalized Poland Scheraga 
(GPS) model in the grand canonical ensemble, in which the partition 
function $Z^{K}(z_{1},z_{2})$ depends on the fugacities, $z_{1}$ and 
$z_{2}$, of, respectively, intact and broken base pairs as:
\begin{equation}
Z^{K}(z_{1},z_{2}) = \sum_{N_{1}, 
N_{2}}z_{1}^{N_{1}}z_{2}^{N_{2}}Z^{K}_{N_{1},N_{2}}
	\label{gpf}
\end{equation}
The partition function, $Z_{N_{1},N_{2}}^{K}$ of a chain with 
prescribed numbers, $N_{1,2}$ of respectively intact and broken base 
pairs is related to the corresponding partition function 
$Z_{N_{1},N_{2}}^{K=0}$ of the stress-free PS model by
\begin{equation}
Z^{K}_{N_{1},N_{2}} = Z^{K=0}_{N_{1},N_{2}}\exp \left[ - \beta 
K\frac{\left(N_{2}- \sigma \left(N_{1}+N_{2}\right) 
\right)^{2}}{N_{1}}\right]
	\label{newz}
\end{equation}

Applying Cauchy's Theorem to Eq.  (\ref{gpf}), we obtain
\begin{equation}
Z^{K}_{N_{1},N_{2}} = \left(\frac{1}{2 \pi i}\right)^{2}\oint 
dz_{1}\oint dz_{2} \frac{e^{- \beta K 
N_{2}^{2}/N_{1}}Z^{K=0}(z_{1},z_{2})}{z_{1}^{N_{1}+1}z_{2}^{N_{2}+1}}
	\label{contint1}
\end{equation}
The functional form of $Z^{K=0}(z_{1},z_{2})$ near the thermodynamic 
singularity of the PS models adopts one of two different forms:
\begin{mathletters}
\begin{eqnarray}
Z^{K=0}(z_{1},z_{2})& \propto  &
\left((z_{e}(I)-z_{1}) + c(I)t +b(I)(z_{e}(I) - z_{2}) 
-a(I)(z_{e}(I)-z_{2})^{p_{I}}\right)^{-1} \label{1order} \\
Z^{K=0}(z_{1},z_{2})& \propto &\left((z_{e}(II) -z_{1}) + c(II)t 
+a(II)(z_{e}(II)-z_{2} )^{p_{II}}\right) ^{-1} \label{2order} 
	\label{twoorders}
\end{eqnarray}
\label{orders}
\end{mathletters}
where $z_{e}(I,II)$, $c(I,II)$, $a(I,II)$ and $b(I,II)$ are positive 
numbers that depend on $\beta \epsilon_{0}$.  The parameter $t$ is the 
reduced temperature, which vanishes at the melting point.  The 
exponents $p_{I}>1$ and $0<p_{II}<1$ play a key role in the 
thermodynamic properties of the melting transition.

Case I (Eq.  (\ref{1order})) corresponds to PS models in which 
excluded volume interactions are included, both for the more flexible 
single strands and for double strands.  These interactions lead to a 
first order melting transition.  However, in case II (Eq.  
(\ref{2order})) excluded volume interactions are included only between 
single strands.  In Case II, at the critical temperature $t=0$ a pole 
in the complex plane merges with the branch cut starting at 
$z_{e}(II)$, corresponding to a continuous phase transition.  The mean 
length of a denaturation bubble diverges at $t=0$, while the 
correlation length, $\xi(t)$, diverges as $t^{1/p_{II}}$, and the 
specific heat exponent $\alpha = 2-1/p_{II}$.  The power $p_{II}$ thus 
plays the role of a critical exponent, and hyperscaling is obeyed with 
the dimensionality $d$ equal to one.  In Case I, the mean size of a 
bubble remains finite at the temperature.  The melting transition is 
first order.

Performing the integration over $z_{1}$ in Eq. (\ref{contint1}) we 
obtain
\begin{equation}
Z^{K}_{N_{1},N_{2}} \propto \oint dz\exp \left[ -Nf_{I,II}(z,M) 
\right]
	\label{contint2}
\end{equation}
with $M \equiv N_{2}/N$ the fraction of all pairs that are broken, and 
$z=z_{2}$.  The function $f_{I,II}$ is the sum of entropic and 
enthalpic terms:
\begin{equation}
f_{I,II}(z,M) = M \ln z + (1-M) \ln \left(Z_{I,II}^{K=0}(0,z) 
\right) + \beta K(M-\sigma)^{2}/(1-M)
	\label{fdef}
\end{equation}
In the thermodynamic limit $N \rightarrow \infty$, but with $M$ finite, 
the partition function is dominated by the minimum of $f_{I,II}$ with 
respect to both $z$ and $M$.  Effectively $f_{I,II}(z,M)$ plays the 
role of a Landau variational free energy per base pair.  This leads to 
the following coupled equations of state for $M(t)$:
\begin{mathletters}
\begin{eqnarray}
\ln \left\{ \frac{Z^{K=0}(0,z)}{z} \right\} & = & \beta K 
\frac{(M-\sigma)(2-M-\sigma)}{(1-M)^{2}}  \label{eos1} \\
\frac{M}{1-M} & = & -\frac{z}{Z^{K=0}(0,z)}\frac{\partial}{\partial z 
} Z^{K=0}(0,z)
\label{eos2}
\end{eqnarray}
\end{mathletters}

\begin{figure}
\centerline{\epsfig{file=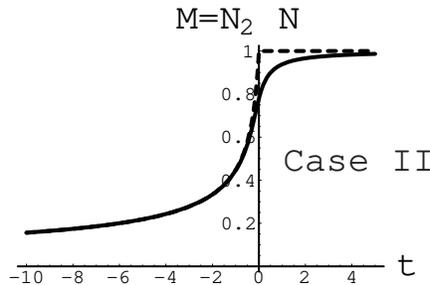,height=1.5in}}
\caption{Dependence of the fraction of broken pairs, $M=N_{2}/N$ on 
the reduced temperature, $t$, for Case II with $p_{II}=0.5$ and 
$\sigma=0$.  The dashed curve shows $M(t)$ for the $K=0$ case, i.e.  
with denaturation-induced twist energy fully relaxed.  The solid curve 
shows $M(t)$ for $\beta K =1$.  In the second case there is no 
thermodynamic singularity.  The melting transition in the 
strain-unaffected case is at $t=0$.}
\label{fig:cont}
\end{figure}

\begin{figure}
\centerline{\epsfig{file=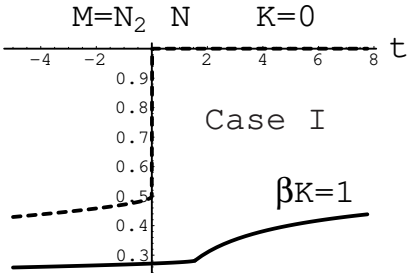,height=1.5in}}
\caption{Dependence of the fraction of broken pairs, $M=N_{2}/N$, on 
the reduced temperature, $t$, for Case I with $p_{I}=1.5$. The 
dashed curve shows $M(t)$ for $K=0$, and the solid curve shows $M(t)$ 
for $\beta K =1$. In the second case, there is a third order 
thermodynamic instability.}
\label{fig:firstorder}
\end{figure}

We first focus on the case $\sigma =0$.  Using Eq.  (\ref{2order}) for 
$Z^{K=0}(0,z)$, we obtain the results for $M(t)$ shown in Fig.  
\ref{fig:cont}.  For Case II \emph{there is no thermodynamic 
singularity at $t=0$}.  The function $M(t)$ has a smooth, sigmoidal 
behavior, even for very small values of $K$, and there is no specific 
heat anomaly.  Physically, this is due to the continuous arrest of the 
denaturation by the build-up of strain energy as more and more bubbles 
open.  The transition is never completed, and the ``denatured'' case 
consists of large bubbles connected by tightly twisted double-stranded 
sections.  For Case I, the thermodynamic singularity is not completely 
removed, but the first order phase transition is transformed into a 
weak, third order transition \cite{Tisza} (see Figure 
\ref{fig:firstorder}).

\begin{figure}
\centerline{\epsfig{file=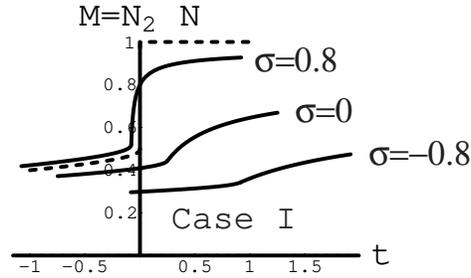,height=1.5in}}
\caption{Dependence of the fraction of broken pairs, $M=N_{2}/N$, on 
reduced temperature, $t$, for different values of the degree of over- 
and underwinding for Case I with $p_{I} = 1.5$ and $\beta K =1$. For 
$\sigma \neq 1$ there is a third-order thermodynamic singularity. For 
$\sigma =1$ the first-order singularity of standard Case I PS models is 
recovered. The dashed curve is of the fraction of broken pairs when 
$K=0$. }
\label{fig:u3}
\end{figure}

We now turn to the the effects of non-zero values for the underwinding 
parameter $\sigma$ on the melting transition.  In the case of a 
transition that is first order in the absence of strain-related 
effects (Case I), we find that, with one exception to be outlined 
below, underwinding or overwinding do not change the qualitative 
effect of strain energy on the transition.  The ``renormalization'' of 
the order of the transition is the same, although the transition 
temperature will change with $\sigma$.  Figure \ref{fig:u3} displays 
the dependence of the denatured ratio, $N_{2}/N$ as a function of 
reduced temperature for different values of $\sigma$.  Of note is the 
fact that underwinding can lower the transition temperature 
\emph{below} its value for $K=0$.  Underwinding also enhances 
denaturation at low temperatures, even though the full transition is 
frustrated.  This qualitative result is in general accord with recent 
experimental observations \cite{viglasky}.

\begin{figure}
\centerline{\epsfig{file=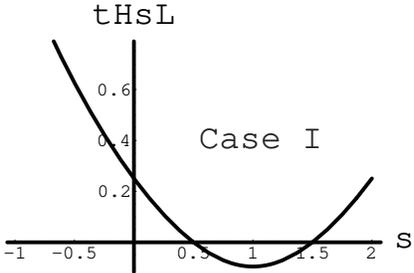,height=1.5in}}
\caption{A plot of the transition temperature, $t(\sigma)$, of under- 
and overwound DNA as a function of the degree, $\sigma$, of over- or 
underwinding, for Case I wilth $p_{I}=1.5$ and $\beta K =1$.  When 
$\sigma=1$, the interwinding of the two strands of duplexed DNA has 
been removed.}
\label{fig:trans}
\end{figure}

Figure \ref{fig:trans} displays the transition temperature of under- 
and overwound DNA as a function of $\sigma$.  Note the minimum in the 
transition temperature at $\sigma =1$.  At the special value $\sigma 
=1$, which corresponds to a strand of duplexed DNA that has been 
underwound to the point that the two individual strands are completely 
unwound, the transition temperature is lowered, and the first order 
transition to full denaturing is recovered.

It thus seems clear that inclusion of denaturing-induced stresses in 
the PS model for circular DNA either removes the thermodynamic 
singularity (Case II) or weakens it to such an extent that one would 
encounter great difficulty in observing it experimentally (Case I).  
The one exception is the introduction of ``complete'' underwinding, 
corresponding to underwinding parameter $\sigma=1$.  However, an 
important limitation on the present model is that we have assumed that 
the strains created as the result of the appearance of denaturation 
bubbles are taken up entirely in the form of twist.  It is well-known 
that torsional stess on undenatured portions of DNA also produces 
\emph{writhing}, e.  g., formation of heterogeneous three-dimensional 
interwound plectonemic structures.  It is likely that the effect of 
writhing will be to suppress the re-appearance of the first order 
transition for $\sigma =1$, because plectonemic structures are 
inherently inimical to strand separation.  Such structures might 
account for the non-monotonicity of melting temperature as a function 
of $\sigma$ observed in \cite{viglasky}.

\end{document}